\documentclass[useAMS,usenatbib,usegraphicx]{mn2e}

\title[Accretion in Mira Winds]{Continuous and Burst-like
 Accretion\ onto Substellar Companions in Mira Winds}

\author[C. Struck, B. E. Cohanim, and L. A. Willson] {Curtis
Struck$^{1}$\thanks{E-mail: curt@iastate (CS); spyder8@mit.edu (BEC);
lwillson@iastate.edu (LAW)}, Babak E. Cohanim$^{2}$\footnotemark[1]
and Lee Anne Willson$^{1}$\footnotemark[1]\\
$^{1}$Dept. of Physics and Astronomy, Iowa State University, Ames, IA
50014 USA\\
$^{2}$Center for Space Research, M.I.T., Cambridge, MA 02139 USA}

\begin{document}


\pagerange{\pageref{firstpage}--\pageref{lastpage}} \pubyear{2002}

\maketitle

\label{firstpage}

\begin{abstract}

We present numerical hydrodynamical modeling of the effects of a giant
planet or brown dwarf companion orbiting within the extended
atmosphere and wind formation zone of an approximately solar-mass Mira
variable star. The large-scale, time-dependent accretion flows within
the radially oscillating and outflowing circumstellar gas around Miras
are related to Bondi-Hoyle-Lyttleton flows, but have not, to our
knowledge, been previously modelled. The new models presented in this
paper illustrate the changes in accretion and wake dynamics as the
companion mass is varied over a range from 10 to 50 Jupiter masses
($M_J$), and generalize the results of the single model
we presented in an earlier paper.

The character of the accretion onto the companion changes greatly as
the companion mass is increased. At the lowest companion masses
considered here, a low continuous rate of mass accretion is punctuated
by large, nearly periodic bursts of accretion. When the companion mass
is large, the mass accretion has both a continuous part, and a rapidly
varying, nearly stochastic part. Surprisingly, the angular momentum of
the accreted gas shows an opposite trend with mass, varying nearly
periodically at large companion masses, and stochastically at low
masses. These trends can be understood as the result of the interplay
between the shocks and radial oscillations in the circumstellar gas,
and the wake flow behind the companion. Boundary conditions also
affect the character of the accretion. The equation of state, however,
is found to have little effect, at least for gamma-law gases, with
gamma in the range from 1 to 5/3.

Models with accretion bursts may produce observable optical
brightenings, and may affect SiO maser emission, as we suggested in
previous papers. Interruptions of continuous accretion, or shadowing
effects, could give rise to bursts of dimming in the optical. Such
dimming effects are likely to be correlated with bursts and optical
flashes, helping to explain some rather mysterious Hipparcos
observations.

\end{abstract}


\section{Introduction}

The number of known extrasolar planets orbiting nearby stars is now
more than 100 (see e.g., the website: http:\\exoplanets.org). Because
of strong observational selection effects most of the systems
discovered first consisted of a Jovian planet in a very close orbit
around the parent star. Multiple planet systems have been discovered
now, as have systems with Jovian planets or low mass brown dwarfs with
orbital radii more like those of our own solar system.

The question of how these diverse solar systems formed has generated
much interest. As in our own solar system, there are also many
interesting questions about the fate of these systems when their
parent stars reach the end of their evolution. However, it is
difficult to discover planets around luminous, bloated giant
stars. Thus, if this study is to advance in the near future, unique
and strong observational signatures of planets around evolved stars
must be found.

\citet{sok96a} and \citet{sok00} (also see \citealt{mas98} and
\citealt{mas99}) have suggested that the elliptical, bipolar and even
non-axisymmetric shapes of planetary nebulae could be the result of
their binary nature, with the binary companion being a substellar
object in many cases. \citet{sok96b} has also proposed that jets in
planetary nebulae could be formed as a result of the engulfment of
giant planets or brown dwarfs in the envelope of AGB
stars. \citet{liv02} recently proposed that such engulfments could
modify the mass loss of the AGB star. They estimated that at least
3.5\% of solar type stars would be affected. These phenomena provide
strong observational signatures, but the disadvantage is that the
substellar companion must disappear in most cases (and in a short
time) to produce them.

We have suggested that other observable signatures may result from the
accretion of AGB wind or extended atmosphere material onto closely
orbiting giant planet or brown dwarf companions. These include
enhancing SiO maser emission in Mira variables \citep{str88}, and
producing short time-scale optical variations \citep{wil01}, such as
reported in Hipparcos observations of Miras \citep{del98}. Previously
(\citealt{str02}, henceforth Paper I) we presented some preliminary
numerical hydrodynamical models of the accretion process.

These signatures do not require the companion to be swallowed, but the
optical variations are more subtle and difficult to detect. The
masers, in particular, may have multiple causes, so the role of
companions is difficult to decipher. Moreover, the companion masses
and radii are restricted by the need to have a strong enough accretion
interaction to produce the observable phenomena.

When solar-type stars evolve to the AGB stage, their outer atmospheres
expand out to nearly an astronomical unit. During parts of this
evolutionary stage they experience radial pulsations as Mira or
semi-regular variable stars, suffer thermal pulses, and generate winds
that increase with time. The models of \citet{bow88} show how
pulsational shocks steepen as they run down the pressure gradient
outside the photosphere, utimately lifting material into an
oscillating extended atmosphere. Further out, where gravitational
periods are longer, individual gas elements are not able to complete a
cycle between shocks, and a wind develops \citep{hil79}. Radiation
pressure on dust grains also contributes to driving the wind (e.g.,
\citealt{bow88}, \citealt{gai99}, \citealt{wil00}). The dominant
pulsational modes in Miras are radial, so it is likely that the wind
is spherically symmetric, unless the atmosphere has been spun up or
otherwise modified by interaction with a close companion. The
discovery of concentric shells around some young planetary nebulae
provides strong evidence for circular symmetry in the winds (e.g.,
\citealt{bal01} and references therein).

Any companions within the large optical photosphere of the Mira star
will be swallowed - that is, heated and dispersed as they are dragged
inward. Companions orbiting at distances of greater than about 10 AU
from the central star will accrete, but accretion from the quite
steady and relatively low-density wind at these radii is unlikely to
generate any very strong observational signatures. (That is, unless
the companion is of stellar mass, see \citealt{mas98} and
\citealt{mas99}.)

The best place to produce time-varying observable effects with
accretion onto low-mass companions is closer to the star, where they
can interact with pulsational shocks in the extended
atmosphere. According to the \citet{bow88} models, pulsational shocks,
and quasi-ballistic motions of gas elements with radial velocity sign
changes, persist out to radii of about $10 AU$ in typical Miras. In
\citet{wil01} and Paper I we found, from energy estimates and
preliminary models, that even this close to the central star, it is
unlikely that steady accretion onto giant planets and low-mass brown
dwarfs would produce observable effects. In this region the
pulsational shocks modulate the amount of material that sweeps into
the companion's sphere of influence, though not by enough to give rise
to large emission bursts.

Captured material has little specific angular momentum relative to the
companion, so we expect it to settle into a disc of radius $R$ much
less than the capture radius $R_{cap}$. (For clarity, we will refer to
radii or distances from the companion with a capital letter R, and
distances from the Mira with the lower case r.)  Because the
surrounding medium is oscillating in radius, the sign of the accreted
angular momentum is expected to change periodically. (As discussed
below, the situation is in fact more complicated than this.) Thus, we
expect a disc to build up near the companion with the gas orbiting in
one sense, but to be disrupted when gas orbiting in the opposite sense
begins to rain down. The resulting angular momentum cancellation would
drive infall onto the planet, with a burst-like energy release. The
transient disc essentially serves as an energy reservoir, allowing
substantial burst-like releases. The models presented below do not
spatially resolve the accretion disc, but they can provide some
information on its development.

We note in passing that, since we are only considering companions that
are much less massive than the star, the accretion process is not
Roche overflow, but rather direct hydrodynamic capture from the
surrounding medium. Similarly, estimates of the tidal circularization
or in-spiral time (see Section 2 of \citealt{sok96a} and references
therein), show that tidal effects will only be important over the
brief Mira lifetime (about $3 \times 10^5 yrs.$) if the companion
orbits within about $r = 3.0$ stellar radii of the star. The
oscillating extended atmosphere extends much farther out than
this. The orbital radius adopted in the models below is 3.2 times the
radius of a solar mass Mira (e.g., \citealt{bow88}). (There is
controversy in the literature about Mira radii, but larger Mira radii
would also imply larger oscillatory zones, and not change the
conclusion that tidal effects are only important in the lower part of
this zone.)

The bow shock, wake and accretion phenomena in this problem are
examples of the Bondi-Hoyle-Lyttleton class of wake accretion flows
(\citealt{hoy39}, \citealt{bon44}, \citealt{cow77}). In the last 20
years there has been a great deal of study of the problem of accretion
in X-ray binaries, where a compact object orbits within the wind of a
massive star (e.g., \citealt{fry87}, \citealt{fry88}, \citealt{taa88,
taa89}, \citealt{ho89}, \citealt{sok91}, \citealt{taa91},
\citealt{liv91}, \citealt{ish93}, \citealt{ruf94, ruf95, ruf97,
ruf99}, \citealt{ruf94b}, \citealt{ruf95b}, \citealt{fog97, fog99}, \&
\citealt{pog00}). These flows are very similar to those of the present
work - unstable, oscillating wakes and bursty accretion of low angular
momentum gas are also common in them.  We will make specific
comparisons to this work below. However, we note that because the
time-dependent hydrodynamics is driven by ``external'' pulsational
forcing in the present problem, it is fundamentally different than
that of the flip-flop internal instability found in models of X-ray
binary stars.

In the earlier work we also found that companions must be considerably
more massive than Jupiter to produce observable bursts, i.e., at
least 10 times as massive. The brown dwarf/planet boundary is usually
defined as the critical mass for deuterium burning, about 13 $M_J$
(Jupiter masses). In the models below we consider companion masses in
the range 10-50 $M_J$. As the companion mass is increased through this
range we find an interesting transition in the wake and accretion
hydrodynamics. If we further increase the companion mass much above
this, e.g., $M \ge 80 M_J$ the companion would no longer be
substellar, and we would be entering the domain of the symbiotic star
phenomenon, which is beyond the scope of this paper.

Once we have restricted consideration to systems with brown dwarf
companions orbiting within 10 AU of a Mira variable, we must return to
the question of how common are such systems? The question is
especially important in light of the two points: 1) most of the
recently discovered brown dwarfs are young, freely floating in
star-forming regions (e.g., \citealt{rei99}, \citealt{luc00},
\citealt{zap00}), and 2) for several years after the discovery of
extrasolar planets, it appeared that there was a ``brown dwarf
desert.''  The latter point refers to the fact that as extra-solar
planets were being discovered, brown dwarf companions were not found
(see review of \citealt{mar00} and references therein). 

The difficulty posed by this point has been mitigated
recently. Specifically, the first 50 extrasolar planets discovered
included only 3 objects with $M sin i \ge 10 M_J$, but all had orbital
radii much less than $1 AU$. However, the second 50 planets discovered
include 5 objects of about this mass or somewhat greater, and all have
orbital semi-major axes of greater than $1 AU$.  Thus, near the
planet/brown dwarf boundary, the desert is less deserted. 

Some specific systems are also noteworthy. These include the HR 7672
(HD 190406) system, with a solar analog star. The mass of its
companion is estimated to be in the range of about $50-80 M_J$, and
its orbital semi-major axis is about 14 AU \citep{liu02}. It provides
an example of a high mass brown dwarf companion of a solar type
star. The more massive of the two companions in the HD 168443 system
has $M sin i = 17.2 M_J$, and a semi-major axis of 2.9 AU
(eccentricity of e = 0.20) \citep{mar01}, which very much fits the
range of this study. $\iota$ Draconis (HD 137759) is a roughly solar
mass K giant star with a companion minimum mass of $8.9 M_J$, and
whose semi-major axis is about 1.3 AU (eccentricity 0.7, see
\citealt{fri02}). This planet, the first discovered orbiting a giant,
may be doomed when the parent reaches the AGB stage.

These examples show that while systems with closely orbiting brown
dwarf companions may be quite rare, they do exist. In the following
sections we investigate the accretion hydrodynamics they generate at
the end of the parent stars' life, and the consequences for
observation.

\section[]{The Numerical Models}

The models were produced with a smoothed particle hydrodynamics code,
which is described in Paper I, and \citet{str97}. Since the code is
documented in these papers, here we will only describe the initial
conditions, boundary conditions, and scaling issues relevant to the
present models.

First of all, the gravities of the parent star and companion were
approximated as softened point masses. The SPH smoothing length in
these models was fixed in space and time to a value of 0.1 length
units. The unit of length is taken to be 1.0 AU, and the time unit is
1.0 yr. The gravitational softening length of the companion was set to
0.08 length units. 

Most of the runs described below used 157,500 gas particles
distributed in a complete ($360^{\circ}$) annulus in the orbital plane
of the companion.  Initially, the particles were placed on a circular
grid with constant radial separations between particle circles. A
fixed number of particles were placed on each circle, since this gives
a $1/r$ density profile in two dimensions, which is like the observed
$1/r^2$ profile in three dimensions. The particles were also given a
slight random offset from this grid, and a small random velocity (5%
of the sound speed in each of the 2 dimensions), to produce a smooth
distribution. Particle radial and azimuthal velocities were
initialized to zero, except for the random part. The parent simulation
code is three dimensional, but the present simulations are two
dimensional.

In the model presented in Paper I, an isothermal equation of state was
used with an initial temperature of $T \simeq 1700 K$. The simulations
described below also used a polytropic or gamma law equation of state,
$P \propto \rho^{\gamma}$. Runs with different values of $\gamma$ were
carried out to test the effects of this factor. The Mira atmosphere
and wind models of \citet{bow88} show that the presence of a small
amount of dust can so enhance the cooling behind the pulsational
shocks that the extended atmosphere is nearly isothermal. However,
Bowen models with no dust show large temperature variations. Thus, the
isothermal case may be the most relevant, but is not the only case of
interest.

The first boundary condition, for the inner boundary, is a wall that
is moved inward and outward on a cycloidal trajectory, like that of a
ballistic particle at the same distance from the star. The radial
range of this motion is from r = 2.0 to about 2.5 AU, and its period
is about one time unit.

As in the model of Paper I, we neglect the effect of radiation
pressure on grains.  Then, as noted in that paper, pulsational waves
generated from below with the piston-like boundary, and a zero
pressure outer bound, reproduce the main features of the flow in
Bowen's (dust-free) models, including, approximately, the velocity
profile.

In the model of Paper I the outer boundary was an impervious,
reflecting wall, that was moved steadily outward. This motion avoided
reflection shocks from the outer boundary, and was responsible for the
low pressure that allowed the development of a wind at larger
radii. However, since a fixed number of particles were used, the gas
density steadily decreased, so that the model could not be run for a
long time.

In the models presented below we used a fixed outer boundary. When gas
particles reached this outer bound they were removed from the grid and
reinserted just above the lower boundary, at the same azimuth, and
then moved by a small random displacement to smooth any clumps. Their
velocities were reset to values close to that of the inner
piston. Shock reflections were generally small with this boundary
procedure. The procedure does not, however, completely smooth large
density changes (e.g., behind shocks), so there will be some feedback
effects. No growing instabilities resulted from this feedback, but the
gas behind the pulsational shocks did acquire a more complex,
multi-layered appearance. Since the companion was only followed
through one orbit, the bow shock was not affected.

The final boundary condition is applied on a circle of radius R = 0.05
AU (106 Jupiter radii) around the companion. Large numbers of
particles are captured by the companion. As noted in the Introduction,
the particles have quite small values of specifc angular momentum
relative to the companion, so we expect them to fall into a very small
region around the companion. This region cannot be resolved by the
present code, so their fate is determined by the boundary
condition. In the model of Paper I they were transferred to the centre
of the region, assigned the same velocity as the companion, and
thenceforth traveled as a single clump with the planet. We will call
this case the clump condition.

In the models presented below, particles found within the boundary
radius are removed, and reinjected into the simulation at the lower
boundary, just like those passing through the outer boundary. We call
this the vacuum condition.

When the clump condition is used the particles in the clump still
exert a hydrodynamic pressure on particles outside the boundary,
computed according to the usual SPH algorithm. In the Paper I model
this pressure was initially small, but grew in magnitude as the clump
grew. Eventually, this pressure resisted accretion, and particles
accumulated on the circular boundary until their pressure and
gravitational force overwhelmed the resistance. This effect was
responsible for the burst-like accretion described in that
paper. There is no pressure resistence when the vacuum condition is
used, and any burst-like accretion must be due to other causes, see
below.

In reality, pressure on the outer accretion flow from a small
accretion disc is not realistic.  On the other hand, such resistance
might be generated by other means. For example, if the companion has a
strong magnetic field, the magnetosphere would provide magnetic
pressure against the partially ionized accretion flow. Energy release
from accetion onto the companion surface may generate a hot corona,
and coronal pressure around the companion. The clump condition crudely
represents such a effects (for a limited duration). However, the
vacuum condition is more realistic on the relatively large scales
modeled here, and ultimately resisting pressures should be modeled on
the smaller scales where they are generated.

\section{Model Results}
\subsection{Basic Hydrodynamics}

\begin{figure*}
\scalebox{0.8}{ \includegraphics{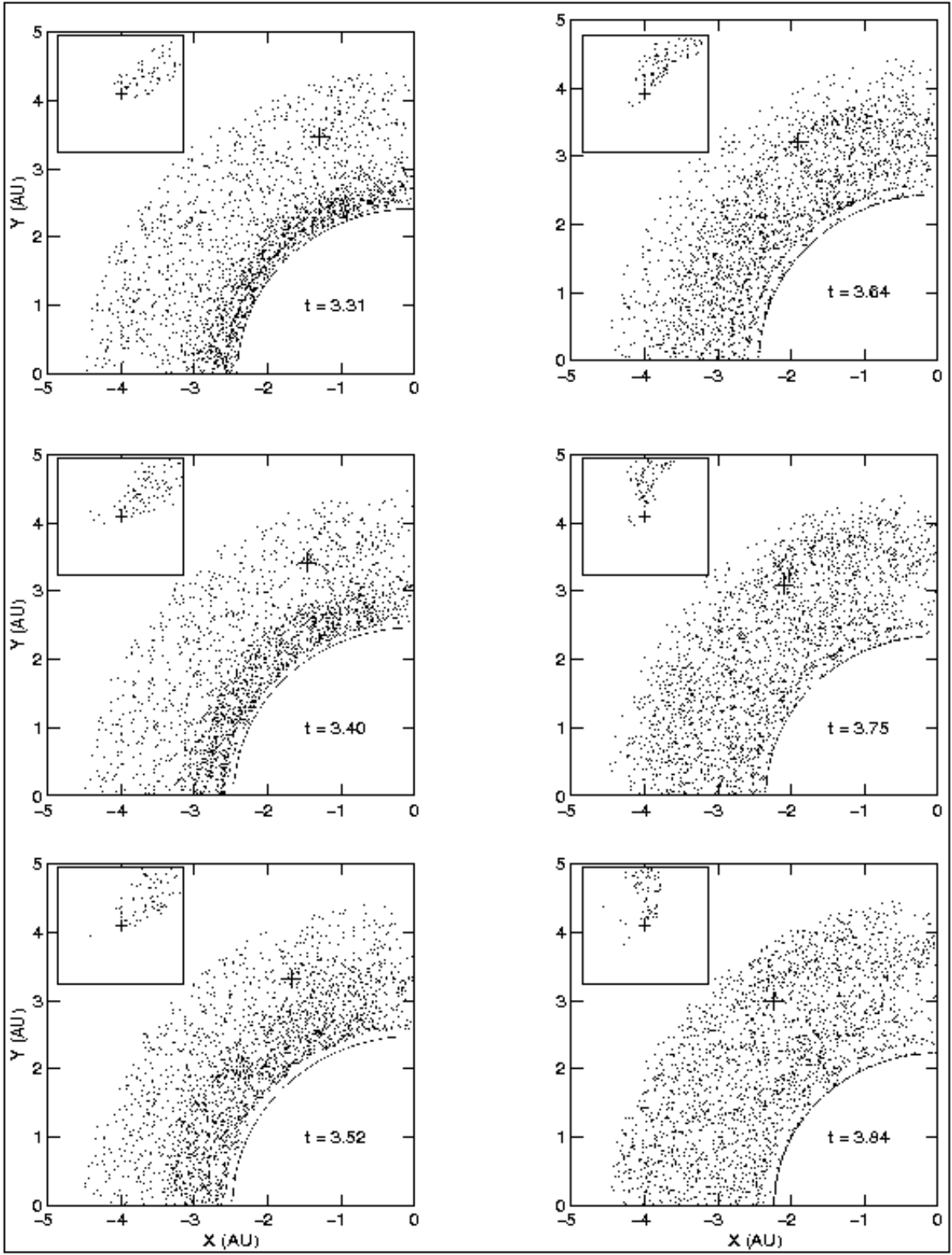}}
 \caption{Wind particle positions in the $30\ M_J$ model at six
timesteps, chosen to cover most of one pulsation cycle. Only a quarter
of the computational grid, the quadrant containing the companion at
these times, is shown. The Mira star is centred at the origin. The
planet's position is marked by a cross in each panel. The size of the
inset is $0.35 AU \times 0.35\ AU$ in all panels. In the insets the
wake is highlighted by plotting only particles with significant
velocities greater than the sound speed (see text for details).}
 \label{f1}
\end{figure*}

We begin each of our model runs with the companion starting on a
counterclockwise orbit from a position near the x-axis, and thus,
moving through the first quadrant.  In a very short time compared to
the orbital period the companion draws gas particles around it into
mildly supersonic motion with it. A classic Bondi-Hoyle-Lyttleton bow
shock and wake structure soon develop (\citealt{hoy39},
\citealt{bon44}). Figure \ref{f1} shows six snapshots through the
course of one Mira pulsation cycle (also see Fig. \ref{f1} of Paper
1), at a time when the companion has moved into the second
quadrant. Particles in the wake and the (incomplete) bow shocks are
highlighted in the insets, which show magnified views of the region
near the companion, and in which only particles with supersonic (i.e.,
greater than 1.0 velocity units) azimuthal velocities are
plotted. Initially the gas particles have zero azimuthal velocities
and only acquire them through interaction with the companion. The
wakes are truncated at the outer boundary, but can clearly be seen for
a length of more than an AU.

Also at the beginning of a model run, the piston at the lower boundary
of the annular grid is started outward. As this and subsequent
pulsational shocks move radially out through the grid they set up a
pattern of radial motions much like those seen in the \citet{bow88}
atmospheric models. We model a minimal radial range needed to capture
this basic kinematics, in contrast to Bowen's models. The snapshots in
Figure \ref{f1} show the progress of one shock from the lower piston
to the outer boundary.

Recycling material at outer boundary back to the inner boundary does
not seem affect the flow much in regions away from the companion. Flow
profiles are quite similar to those of Paper I, where a
moving/reflecting boundary was used.  This is probably because the
pulsational shocks are weak by the time they reach this bound in both
cases. Nonetheless, the shocks driven by the lower piston do acquire
some additional structure from the remnant waves reinserted, but with
some smoothing, from the outer bound, as can be seen in Figure
\ref{f1}. Note that a layer of particles ``stick'' to the lower
boundary (a well known effect in SPH). We inject the recycled
particles just above this layer.

It is clear from the Figure \ref{f1} snapshots and their associated
insets that the radial shocks are able to move the companion
wake. This is reasonable since the wake column density along a radial
line is of the same order as that of the gas immediately behind a
pulsational shock.  The wake is stretched outward when pulsational
shocks sweep over it. The infall in the postshock rarefaction wave
leads to reformation of the wake, and leaves it pointing in a slightly
inward direction. It is also apparent from Figure \ref{f1} that the
wake interaction disturbs the pulsation shock and postshock flow. This
is especially apparent in the fourth and fifth snapshots of Figure
\ref{f1}, where the shock is disturbed over a distance of nearly 2.0
AU behind the companion.

\subsection{Mean Accretion Properties}

In this section we consider the mean properties of the accretion flow
onto the companion, in the next section we consider its more detailed
structure. This accretion is characterized by two radial scales: an
accretion radius (and BHL accretion cylinder) from which it is
gravitationally captured, and an inner scale equal to the radius of
the accretion disc around the planet, where the material settles. As
noted above and in Paper I our models do not resolve the inner
scale. This means that we cannot account for the possibility that some
captured material may be subsequently pushed back out by dynamical
processes at the inner scale, before being accreted onto the
companion. However, we expect this to be a small fraction of the
captured gas.

\begin{figure*}
\scalebox{0.4}{ \includegraphics{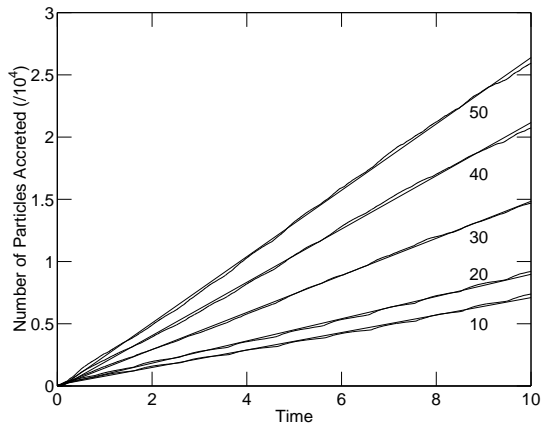}}
 \caption{Cumulative number of particles accreted versus time for the
$10, 20, 30, 40,\ and\ 50\ M_J$ models (labelled). The time unit is one
year, and the y-axis is labelled in units of $10^4$ accreted
particles. See Sec. 4.1 for an estimate of the particle mass. A least
squares mean line is fit to each model curve.}
 \label{f2}
\end{figure*}

In the simulation code, any particle that comes within a radius of
$R_{cap} = 0.05\ AU$ of the companion is assumed to be accreted, and
is removed and recycled. Figure \ref{f2} shows this estimate of the
number of particles captured versus time in five
model runs with companion masses of $10, 20, 30, 40\ and\ 50\
M_J$. The classical BHL accretion radius is,

\begin{equation}
R_{BHL} = \frac{2GM_c}{v_{rel}^2 + c^2},
\end{equation}
where $M_c$ is the companion mass, c is the sound speed in the ambient
gas, and $v_{rel}$ is the relative velocity of the companion and the
ambient gas, i.e., the sum of the companion's orbital velocity and the
current ambient radial velocity (see e.g., \citealt{sha83}). (Note
that the latter velocity is constantly changing in direction and
magnitude, so that the concept of an upstream accretion cylinder out
of which gas is accreted becomes somewhat confused.) The adopted value
of $R_{cap}$ is about equal to $R_{BHL}$ in the $M_c = 10 M_J$ case,
and smaller than $R_{BHL}$ for larger companion masses. (In the limit
that the relative velocity equals the companion orbital velocity,

\begin{equation}
\frac{R_{BHL}}{a} = 0.019\ \frac{M_c}{10M_J} \frac{1.0M_{\odot}}{M_*},
\end{equation}
where a is the companion orbital radius, and $M_*$ the Mira mass.)
Thus, the bow shock and accretion cylinder is resolved only for the
larger mass companions.

The cumulative accretion in each of the models plotted in Figure
\ref{f2} is roughly linear with time. In that figure each curve
derived from the model data is paired with a line derived from a least
squares fit. The deviations between the model data and the fitted line
are small in all cases, though interesting as we will see in the next
subsection. This implies that the accretion is on average constant,
despite the radial pulsations in the extended atmosphere.

\begin{figure*}
\scalebox{0.3}{ \includegraphics{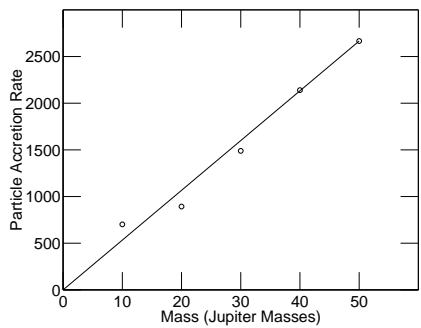}}
 \caption{Mean accretion rate versus companion mass for the $10, 20,
30, 40,\ and\ 50\ M_J$ models. The values on the y-axis are in units
of particles per year, and the derived value for each model is the
slope of the line fit to that model shown in Figure 2. The line
connects the origin to the $50\ M_J$ model, see text for details.}
 \label{f3}
\end{figure*}

Figure \ref{f3} compares the accretion rates of the different
models. Specifically, it shows the average accretion rate over the
duration of each run, plotted against the mass of the companion. In
addition to the data points a line fit is shown. This line simply
connects the best-resolved $50 M_J$ case to the origin, since we
demand zero accretion for a zero mass companion. Most of the other
models lie close to this line except for the poorly resolved $10 M_J$
model.

As in the classic BHL problem most of the accretion in the more
massive companions comes through the wake. In the BHL problem the
accretion radius increases with the companion mass, and the accretion
rate should increase with the square of this radius. That is, with the
square of the companion mass. The goodness of the linear fit in Figure
\ref{f3} appears to represent the two-dimensional BHL result.

However, in this problem, the near linear fit may be more generally
applicable. The oscillatory motions of the ambient gas deliver
material to the companion from a greater range of radial distances
than $2R_{BHL}$. The companion effectively has a greater reach in the
radial direction, though this ``reach'' still seems to scale with the
companion mass. The increase of the size of the accretion cylinder in
the direction perpendicular to the orbital plane with increasing mass
(i.e., the increase in $R_{BHL}$) is smaller and contributes less to
the accretion increase than the increase in the radial direction. If
this is correct, then in three dimensions, the accretion rate will
continue to scale more linearly than quadratically with mass through
the brown dwarf mass range. High resolution, three-dimensional
simulations are needed to resolve this question.

It is also of interest to compare the numerical mass accretion rates
to BHL rates.  E.g., the mean mass accretion rate for a $30 M_J$
companion is about $1500\ particles\ per\ yr$, though we will see
below that burst rates can be much higher. If we adopt a value for the
particle mass of $m_p = 2.0 \times 10^{21}\ g$, as in paper I
(adjusted for the different particle numbers used here, and assuming a
vertical capture distance of $R_{BHL} \simeq 0.2 AU$), then we get the
following conversions:

\begin{displaymath}
1500\ particles\ per\ yr\ = 9.8 \times 10^{16}\ g/s\ = 1.5 \times
10^{-9}\ M_{\odot}/yr.
\end{displaymath}
The stellar mass loss rate is assumed to be of order $2 \times
10^{-7}\ M_{\odot}/yr$. (This is the mass loss rate of the standard
dust-free model of \citet{bow88}, and that used in \citet{hum02},
though it is about an order of magnitude less than the typical Mira
mass loss rate.)  If we compute the Bondi rate,

\begin{equation}
{\dot M} = 4{\pi}{\tilde \lambda} (GM_c)^2 (v_{rel}^2 + c^2)^{-1.5}
\rho, 
\end{equation}
using the average circumstellar gas density $\rho = 2.0 \times
10^{-15}\ g/cm^3$, an efficiency of ${\tilde \lambda} = 1.0$, the
sound speed at the companion's radius ($c \simeq 3.3 km/s$), and set
the relative velocity of the companion equal to its orbital velocity
we get ${\dot M} = 1.6 \times 10^{-9} M_{\odot}/yr$. Thus, the
numerical rate nearly equals the Bondi rate. This seems somewhat high,
yet it is comparable to those \citet{ben97} and
\citet{ruf95b}. However, if the BHL scaling with mass does not hold in
three-dimensions, then accretion rates in this problem will be very
different than BHL rates at the extremes of the brown dwarf mass
range.

Finally, we note that the $30\ M_J$ run was repeated with several
different values of $\gamma$, the exponent in the equation of
state. We found hardly any change in the accretion rate as $\gamma$
was varied. The analytic interpolation formula of \citet{fog97} also
shows little deviation from the BHL accretion rates at Mach numbers
greater than a few, when $\gamma$ is varied from 4/3 to 5/3. Numerical
studies of the flip-flop instability indicate that there is no
instability at $\gamma = 1.0$, so the accretion is steady (see
\citealt{ruf99}, Sec. 3.2). This contrasts greatly with the present
externally driven case.

\subsection{Time-Dependent Accretion}

\begin{figure*}
\scalebox{0.8}{ \includegraphics{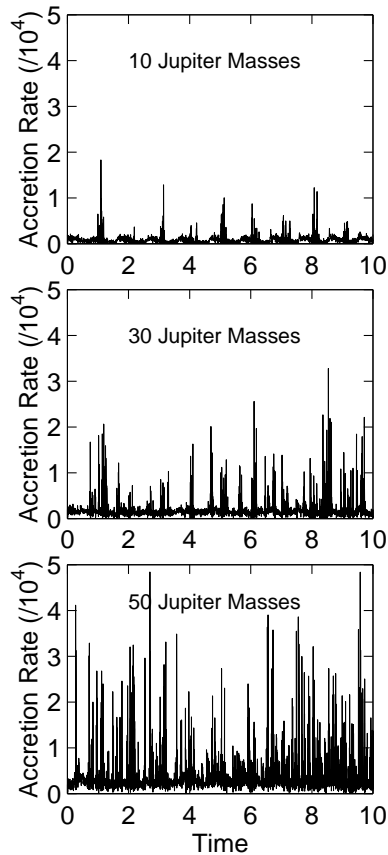}}
 \caption{Comparison of the instantaneous particle accretion rate
versus time for the $10, 30,\ and\ 50\ M_J$ models, with separate
panels (labelled) for each model. The y-axis is labelled in units of
$10^4$ accreted particles.}
 \label{f4}
\end{figure*}

In this section we consider time-dependent aspects of the accretion
more carefully. We begin with Figure \ref{f4}, which shows the
instantaneous accretion rate (${\Delta}M/{\Delta}t$) in three of the
models. We note the presence of sharp, high amplitude peaks in the
accretion rate. There are also sharp decreases relative to the small,
but nonzero, mean accretion rate. In all cases the accretion bursts
are clustered. The cumulative effect of these clusters is presumably
responsible for the deviation from the mean lines in the curves of
Figure \ref{f2}.

A second feature of Figure \ref{f4} is the fact that the burst
clusters are nearly periodic at low companion mass, but appear to
become progressively more random, and frequent, as the companion mass
increases. At low masses, the burst period is about equal to the
pulsation period, and in fact, the bursts are coincident with
pulsational shocks overrunning the companion. It is natural that the
companion should accrete more from dense post-shock layers (which is
also when the radial velocity changes sign). The bursts are more
random with high mass companions because, as we will describe below,
of the processing of this material through the longer wake.

Before proceeding we note the similarity of some of these features to
those of simulations of flip-flop accretion in X-ray
binaries. Instability and quasi-periodic bursts are certainly a
characteristic of accretion in this problem as well (e.g.,
\citealt{taa88}, \citealt{fry88}, \citealt{taa89}, \citealt{liv91},
\citealt{sok91}, \citealt{taa91}, \citealt{ish93}, \citealt{ruf95b},
\citealt{ruf97}, \citealt{ben97}, \citealt{ruf99}, \citealt{pog00}).
Since the flip-flop instability is self-excited, much of the
discussion of this literature has focussed on the question of what
circumstances are required for its development. In particular, issues
such as the effect of pressure and velocity gradients across the Mach
cone have been considered in some detail. Numerical studies have been
plagued by spatial size and temporal limitations, as well as numerical
resolution problems (see \citealt{ben97}, \citealt{pog00}, and
references therein). The present problem has the advantage of being
driven by strong, external perturbations, so we don't have the
difficulty of modeling the slow growth of small amplitude
perturbations under the internal instability.

As a result of this fundamental difference in the two processes, the
details of the accretion burst phenomenon are also different. In the
published flip-flop simulations the amplitude of accretion bursts
relative to the mean never exceeds a factor of about 5. Figure
\ref{f4} shows much larger burst amplitudes. There also seems to be no
equivalent to the systematic change in the character of the
accretion bursts as companion mass is increased, as shown in Figure
\ref{f4}. However, it appears that nonlinear trends with companion
(e.g., stellar black hole) mass have not yet been investigated in the
flip-flop problem.

\begin{figure*}
\scalebox{0.4}{ \includegraphics{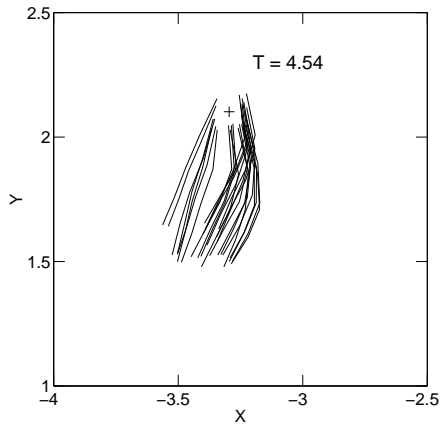}}

 \caption{Partial trajectories of 27 accreted particles relative to
the fixed companion position (marked with a cross) in the $10\ M_J$
model are shown at the time indicated. Part of one quadrant of the
simulation volume is shown. The particle trajectories are constructed
by choosing particles within a capture radius of 0.075 AU at the given
time, and connecting their positions at several previous times with
straight line segments.}

 \label{f5}
\end{figure*}

\begin{figure*}
\scalebox{0.65}{ \includegraphics{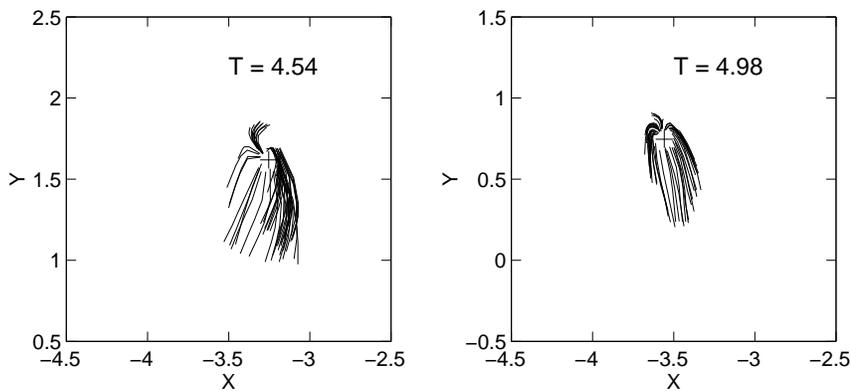}}
 \caption{Partial trajectories of about 100 accreted particles as in
the previous figure, but from the $50\ M_J$ model, and at two
timesteps differing by about half a pulsation period.}
  \label{f6}
\end{figure*}

We noted above that models with larger mass companions draw gas in
from a larger range of radii. Figures \ref{f5} and \ref{f6} provide
many more details about this process for the $10\ M_J$ and $50\ M_J$
models, respectively. These figures show the partial trajectories of
particles that come within a radius of $R_{cut} = 0.075\ AU$ of the
companion, and their subsequent fate. The value of this cutoff radius
is arbitrary, and was chosen to provide an adequate particle
sample. Specifically, once the particles within the cutoff radius were
identified (at the times indicated), their positions relative to the
companion at 4 previous output times were used (with the present
position) to produce the curves. Since the companion position is fixed
in these plots, the primary component of these trajectories is the
relative azimuthal motion toward the companion. (Note: the companion
is located in the second quadrant at these times.)

In Figure \ref{f5} and the first panel of Figure \ref{f6} the time
shown is when a pulsational shock is just about to overrun the
companion. This is evident in the abrupt trajectory changes visible on
the right side of the companion. The second panel of Figure \ref{f6}
shows a time of about half a pulsational cycle after that of the first
panel, when no shock is near the planet. At that time the trajectories
are smoother, and more symmetrically distributed.

Figure \ref{f6} not only shows that many more particles are drawn in
by the $50\ M_J$ companion, but also resolves the capture process
quite well. A large fraction of the accreted particles circle around
the companion, impact the wake, and are drawn down from the inner
wake. The trajectories coming from above (behind in the orbital sense)
are especially interesting. Some particles still escape through the
outer wake, but they are a much smaller fraction with the $50\ M_J$
companion.

With this information, we can now understand the patterns of
instantaneous accretion in the different models. In the low mass
models accretion bursts result from enhanced feeding associated with
the passage of pulsational shocks. Because gas accretes through a
larger wake in the higher mass cases, we would expect longer duration
burst clusters. There is also a possibility of echo bursts following
passage of a pulsational shock, if relatively large amounts of gas are
deposited in the wake, and its accretion is significantly delayed.

The middle panel of Figure \ref{f4}, for the $30\ M_J$ model, does
indeed show wider burst clusters, and bursts appearing at new
phases. When the companion mass is increased to $50\ M_J$ the wake
becomes so massive that the passage of pulsational shocks does not
greately enhance the accretion. There is still some hint of
periodicity in the bursts in the third panel of Figure \ref{f4}, but
the duration of the burst clusters is comparable to the time between
them.

\begin{figure*}
\scalebox{1.0}{ \includegraphics{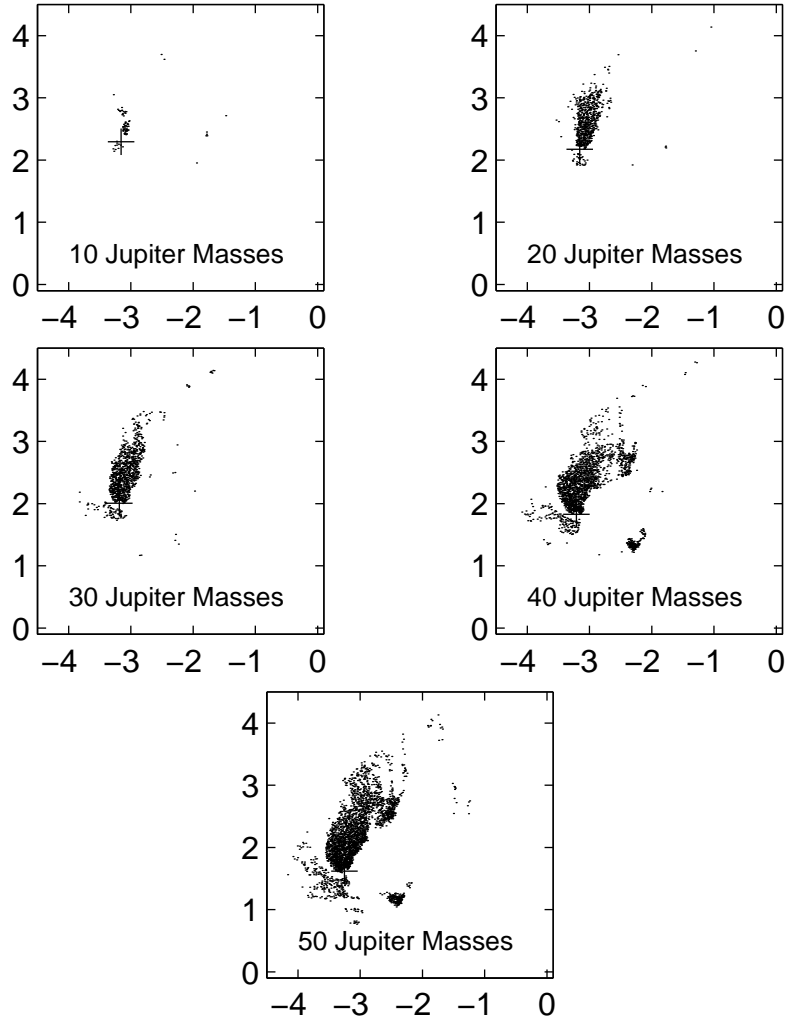}}
 \caption{Positions of all (wake) particles with azimuthal velocities
of greater than 0.8 times the sound speed, with panels for each of the
$10, 20, 30, 40,\ and\ 50\ M_J$ models. }
 \label{f7}
\end{figure*}

To further probe the dependence of the accretion process on companion
mass, it is worth looking at the wakes in more detail. Like the insets
of Figure \ref{f1}, Figure \ref{f7} shows only those gas particles
that have acquired significant azimuthal velocities, though in Figure
\ref{f7} we show the 5 models at about the same time, rather than
snapshots of one model. The wakes are much longer and more massive
when the companion mass is large. In fact, in the case of the most
massive companions the graphs underestimate the extent of the wake,
because wake particles reach the outer bound and are recycled. We note
that the particle clumps seen in the lower centre part of the graphs
for $M = 40, 50\ M_J$ are apparently bow shock particles that have
moved to the outer bound and have been recycled.

Figure \ref{f7} also shows us that the bow shock is barely detectable
in the low mass cases, and asymmetric in the high mass cases. The
phase dependence of this asymmetry can be seen in Figure \ref{f1}. Gas
behind the bow shocks is pushed outward, like wake gas, by the passage
of pulsation shocks. Following these passages, material sweeping past
the inner side of the companion encounters little resistance. Material
moving forward toward the outer side of the companion impacts both
bow shock and wake.

\subsection{Accreted Angular Momentum}

\begin{figure*}
\scalebox{0.8}{ \includegraphics{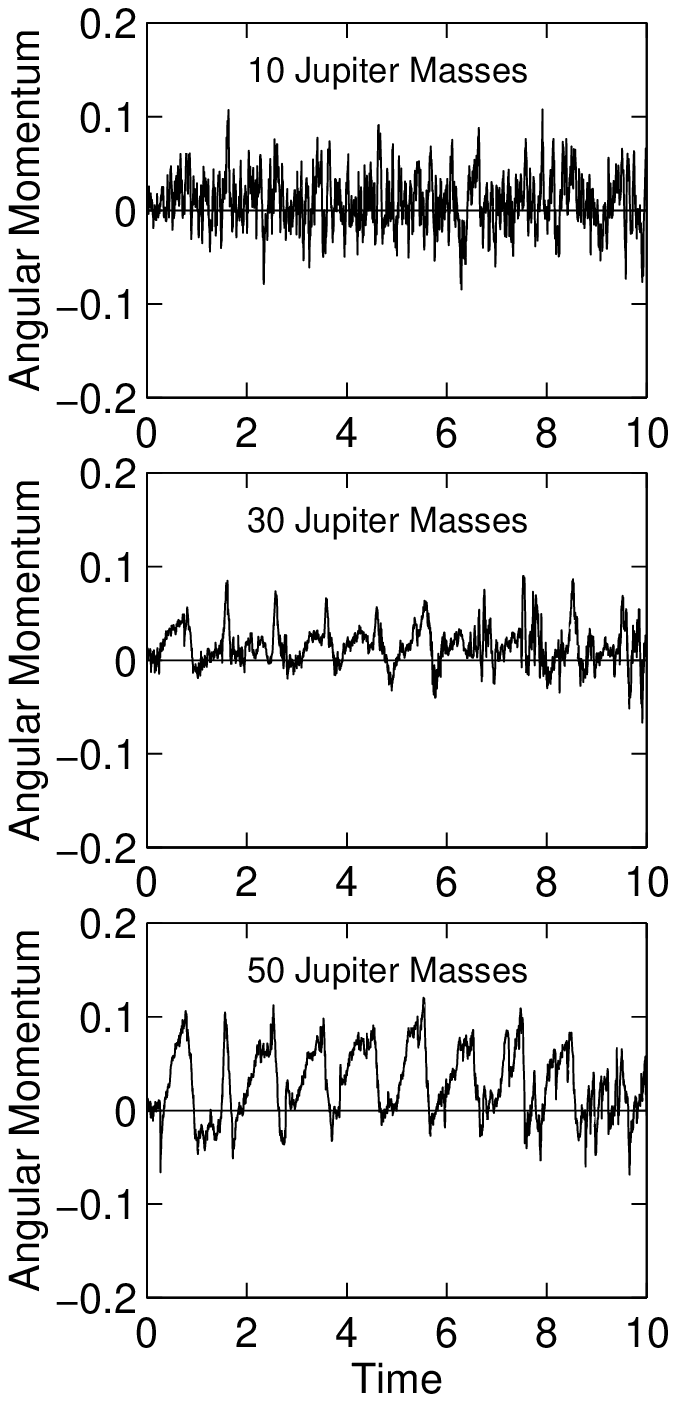}}
 \caption{Comparison of the mean specific angular momentum relative to
the companion versus time for the $10, 30,\ and\ 50\ M_J$ models. The
mean is computed for all particles accreting at that time. The y-axis
is labelled in code units, i.e., ${AU^2}\ {yr^{-1}}$.}
 \label{f8}
\end{figure*}

In Paper I we suggested that the angular momentum vector, relative to
the companion, of material accreted out of the radially oscillating
circumstellar gas would change signs with pulsation phase. Figure
\ref{f8} shows the mean specific angular momentum of the accreted gas
as a function of time in the $10, 30,\ and \ 50\ M_J$ models. This
figure confirms the fact that the accreted angular momentum does have
both positive and negative signs at different times, though the
positive (counterclockwise) sign seems predominant. Moreover, the sign
changes are often quite rapid.

What is even more striking about Figure \ref{f8} is the contrast with
Figure \ref{f4}. It appears that the values of accreted angular
momentum are quite random at low companion mass, where the mass
accretion is periodic, and the angular momentum accretion is periodic
at high companion mass, where the mass accretion is much less
regular. A somewhat more careful reading is that at low mass the
amount of accreted angular momentum is just low. Evidently, even with
the passage of pulsational shocks, gas is accreted relatively
symmetrically in the low mass case, with little net angular
momentum. Some of this effect may be due to low resolution of
asymmetries in the low mass case. However, the size of the accretion
radius is smaller than the thickness of the dense post-shock layer in
the low mass case, and more comparable in the high mass case. Thus,
density asymmetries are not noticeable on the accretion scale in the
low mass case, but are in the high mass case.

\begin{figure*}
\scalebox{0.8}{ \includegraphics{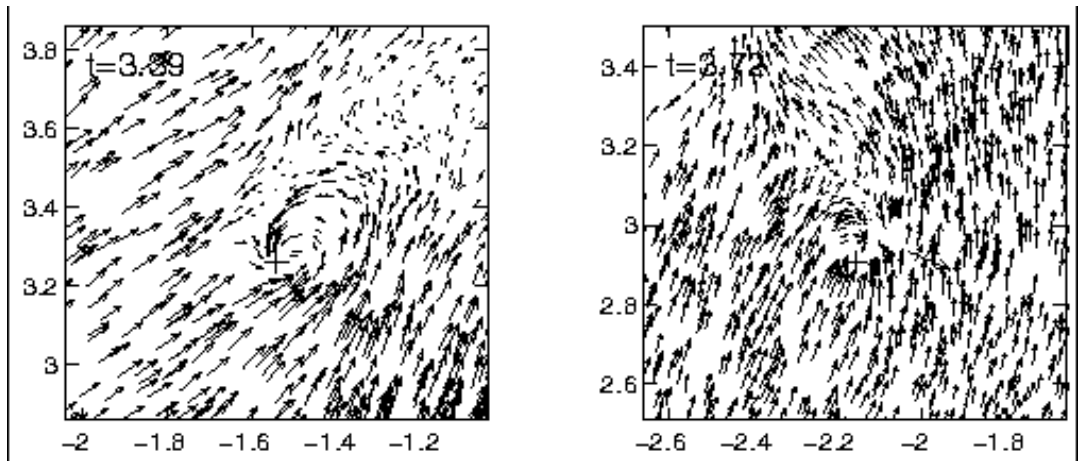}}
 \caption{Particle velocity vectors relative to the companion (cross),
in a small area around the companion in the $50\ M_J$ model, at
t=3.39,  and 3.72. Every fourth particle plotted.}
 \label{f9}
\end{figure*}

The representative particle velocity vectors shown at two times in
Figure \ref{f9} provide more detail on the wake hydrodynamics
responsible for the periodic angular momentum accretion in the $50\
M_J$ model. The right panel shows a time just after a pulsational
shock has passed through the gas, when the velocity vectors have an
outward radial velocity component (toward the upper left), except in
the vicinity of the companion. The left panel shows the velocity field
at a time about midway between shock passages, when most of the
velocities have a radial infall component (toward the lower right).

In the left panel we see gas flowing around the inner side of the
companion, and encountering a shock where the velocity vectors get
smaller and turn. In the post-shock wake this material streams around
and behind the companion, where it encounters a shock separating it
from material flowing in from the outer side of the companion. The
latter stream is turned aside. The vortex likely formed in this sense
because gas coming around on this side encounters the wake and is
shocked. The outer stream hits the wake at a glancing angle an is
deflected. The inner stream finally spirals onto the companion in a
counterclockwise sense.

In the right panel the situation is reversed. It is the outer stream
that impacts the wake, is shocked, and streams around and onto the
companion in a clockwise sense. In this case, when the pulsational
shock has just passed, the wake and its vortex structure is more
compressed. 

Shock passage moves and compresses the wake quite abruptly, accounting
for the rapid drop in accreted angular momentum in the last plot of
Figure \ref{f8}. After that the radial flow slows and reverses, and it
takes most of the cycle to reform the vortex of the left panel of
Figure \ref{f9}. This accounts for the relatively slow rise of angular
momentum in the $50\ M_J$ model.

In sum, the change in the nature of angular momentum accretion from
quite random to nearly periodic is associated with the increasing role
of accretion out of the more distant parts of the wake, and the
formation of wake vortices with alternating senses of rotation. This
is in contrast to the mass accretion, which is periodic in low mass
models because of its immediate sensitivity to the passage of
pulsational shocks.

As in the case of mass accretion, there are both similarities and
differences in the angular momentum accretion of the X-ray binary
flip-flop case and the present case. Like the present case, the value
of the accreted specific angular momentum is highly variable in the 
flip-flop models (e.g., \citealt{ben97}, \citealt{ruf97, ruf99}). The
variations can be of comparable magnitude to those in the present
case, and also include sign changes in the accreted angular
momentum. The variations in the three-dimensional models of
\citet{ruf97, ruf99} are not as regular as the high mass cases shown
in Figure \ref{f8}, but the two-dimensional simulations of
\citet{ben97} are very similar. However, these flip-flop models are
mostly adiabatic, and we recall that there is no flip-flop instability
in the isothermal case. 

\section{Observational Consequences}

In this section we consider the ramifications of the results of the
previous section for observation. Before beginning we recall that the
models above were focussed on a particular region of parameter
space. Specifically, we considered stellar companions in the mass
range of $10-50\ M_J$, orbiting at a radius of $r = 4.0 AU$ from an
approximately $1.0\ M_{\odot}$ Mira, a region in the extended stellar
atmosphere where pulsational motions are still important. In the
discussion below we will also compare the results above to those of
Paper I, and note possible observational effects beyond this parameter
range.

\subsection{Optical Variability}

In \citet{wil01}, in Paper I, and in Section 1 above we considered the
possibility that accretion processes onto a companion might help
explain the short time-scale optical variability in Miras studied by
\citet{del98} in Hipparcos data. In this subsection we reconsider this
issue in light of the new model results. 

The models simulate the flow on the outer or accretion scale, and
provide no direct information about the inner scale, near the
companion. However, the properties of the captured material, like its
time dependent angular momentum, allow us to infer a good deal about
the behavior on the small scale.

Consider first the low mass companions where the accreted angular
momentum varies relatively stochastically, with a low mean value. For
example, in the case of a $10\ M_J$ companion, the specific angular
momentum is about $h = 0.02\ {AU^2}\ {yr^{-1}} = 1.4 \times 10^{17}\
cm^2 s^{-1}$ (see Figure \ref{f8}). If gas elements with this value of
h settle into a circular accretion disc around the companion, without
loss or gain of angular momentum, then the radius of that disc would
be of about,

\begin{equation}
\frac{R}{R_J} = {\biggl(}\frac{h}{9.5 \times 10^{16}}{\biggr)^2} 
{\biggl(}\frac{10\ M_J}{M}{\biggr)^{1/2}},
\end{equation}
where $R_J$ is the radius of Jupiter. In the example with a $10\ M_J$
companion, and $h = 0.02\ {AU^2}{yr^{-1}}$, this equation yields $R =
2.3 R_J$. Thus, at the low end of the mass range considered above, the
significant fraction of the accreted gas with specific angular
momentum somewhat smaller than this value (or companion mass or radius
somewhat larger) can fall directly onto the companion surface. The
rest may be found in a small, and continually disturbed, accretion
disc. (We note that such discs are much smaller than those of
Mastrodemos \& Morris (1998, 1999), whose stellar mass companions,
located in steady outflows, accrete gas with higher values of the
specific angular momentum.)

We discussed mass accretion rates in Section 3.2. With those rates, and
if we assume that accreted particles hit the surface of a $30\ M_J$
companion of radius about $1.0\ R_J$ with
about free-fall velocity, and that essentially all of their
kinetic energy is converted into optical luminosity, then the
magnitude of this luminosity is,

\begin{equation}
L = \frac{1}{2} {\dot M} v_{ff}^2 = 0.090 L_{\odot} {\biggr(}
\frac{v_{ff}}{330\ km/s}{\biggr)^2}.
\end{equation}
This value is for accretion bursts and is correspondingly smaller for
the low accretion rates between bursts. It also assumes an optimistic
conversion of accretion energy to optical photons. On the other hand,
it would be substantially increased for a somewhat more massive
companion in the denser wind of a more evolved Mira. It would be
reduced in the case of greater star/companion separations.

These results are in agreement with those of our earlier papers, where
we found that there would probably not be sufficient energy in the
mean accretion onto giant planets to produce optical flashes. We
conjectured that gas with angular momentum of one sign might be
accreted for about half a pulsation cycle, and stored in a disc.  Then
in the second half of the cycle a comparable mass of gas with angular
momentum of opposite sign might rain down and flush the disc onto the
companion. The large mass of stored gas could produce luminosities
large enough to be observable at stellar minimum light.

In the case of X-ray binaries, this disc storage and flushing has long
been considered as a possible explanation for the rapid changes in the
luminosity and period of X-ray pulsars \citep{taa88}. More recently,
\citet{ben97} estimated that in favorable cases, angular momentum
accretion reversals could be responsible for abrupt spin changes in
some X-ray pulsars.

Figure \ref{f8} suggests that for the lower companion masses there is
no periodicity in the accreted angular momentum, so the simple storage
mechanism will not work. Most of the time the accreted angular
momentum has a positive sign, so most of the time some fraction of
this material adds to the small disturbed disc mentioned
above. Occasionally there are bursts of negative angular momentum
accretion. If these bursts contain enough mass, or their angular
momentum is sufficiently negative, then they might dump enough
material onto the companion to produce an observable burst. In
numerical models such random bursts are partially the product of noise
resulting from small particle numbers. Turbulence or similar effects
would likely generate them in the circumstellar environment.

Another possibility is that pressure from hot gas or magnetic fields
could inhibit the flow to small scales until the accreted mass builds
up to a critical level. The result could be a burst of quite large
mass (and small angular momentum), like those in the model of Paper
I. Interaction with a rotating magnetosphere could also add or
subtract angular momentum to accretion disc gas. The study of such
processes is beyond the scope of this paper.

In the case of large companion masses, Figure \ref{f4} shows larger
mass accretion rates, and Figure \ref{f8} indicates quasi-periodic
angular momentum accretion. Thus, the prospects for storage, followed
by large bursts are much better. According to the estimate of Equation
(1), with h values from Figure \ref{f8}, the typical accretion disc
radii are still small, of order a few times $R_J$. However, with the
deeper gravitational potential, and higher mass accretion rates in
bursts (accumulated in storage), the luminosity estimates from
Equation (2) can exceed a solar luminosity. This is more than enough
to give a detectable signal near minimum light in a large amplitude
Mira, as noted in \citet{wil01}.

In fact, storage may not be required for observable optical effects at
companion masses of about $50\ M_J$ or higher. They might also be
possible at fairly large radii where the ambient gas density is
smaller and where pulsational shocks are weak and there is essentially
no radial motion. In this case the optical enhancement would be
relatively constant, with no bursts, which would make it harder to
distinguish from the Mira minimum luminosity. However, we would expect
variations over the companion orbital period, as the companion is more
or less hidden by the extended Mira atmosphere (unless its orbital
plane coincided with the plane of the sky).

\subsection{A Simple Model of the Inner Accretion Scale}

To clarify the issues raised in the previous paragraphs it is worth
extending the numerical results with a simple semi-analytical model of
the inner accretion disc. We adopt the simplifying assumption that all
particles accreted with negative angular momentum and a number of disc
particles carrying an equal amount of positive angular momentum fall
immediately onto the companion. After such events, the accretion disc
is assumed to consist of all the remaining particles, with each
carrying an equal part of the remaining cumulative angular
momentum. With these approximations the model yields the amount of
infall onto the planet and the angular momentum in the disc as
functions of time.

\begin{figure*}
\scalebox{0.8}{ \includegraphics{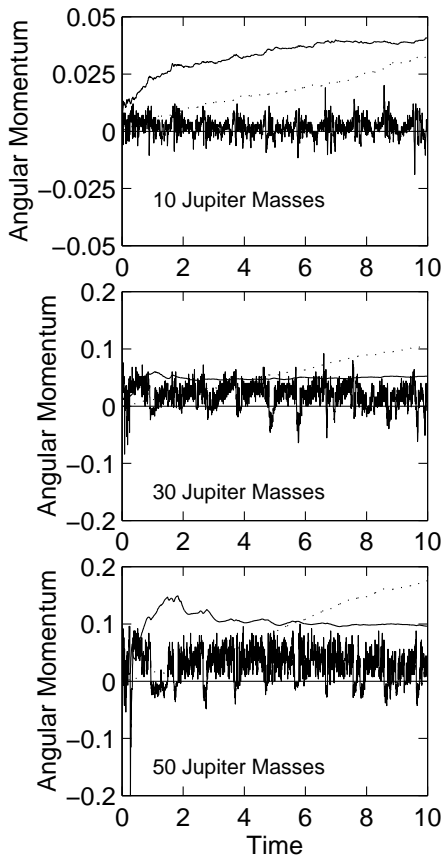}}
 \caption{Comparison of properties of the semi-analytical inner disc
model versus time for the $10, 30,\ and\ 50\ M_J$ models. The rapidly
varying solid curve in each panel shows the amount of mass added to
the inner disc in units of 1000 particles. Positive values indicate
particles added, while negative values indicate particles lost from
the disc that fall onto the companion (see text). The dashed curve
shows the net particle number in the disc in units of $10^5$
particles. The slowly varying solid curve is the net specific angular
momentum of the inner disc in dimensionaless code units, but which has
been multiplied by a factor of 2 in the bottom two plots for clarity.}
 \label{f10}
\end{figure*}

Figure \ref{f10} shows some results of applying this simple model to
data output at every timestep from the numerical hydrodynamical model,
in the cases of the $10,\ 30,\ and \ 50\ M_J$ models. In each case,
the rapidly varying line shows the mass added to the disc versus time,
which is negative if material falls onto the companion.  The smoother
solid curve shows the net specific angular momentum in the disc, and
the dashed curved shows the net particle number in the
disc. Comparison of these particle numbers with those of Figure
\ref{f2} for the outer accretion shows that a significant fraction of
the particles do fall onto the companion in this model.

The values of the specific angular momenta generally seem to settle to
a limiting value in the three cases, though this is not entirely clear
in the $10\ M_J$ model. In the other two cases, this quantity first
peaks, and then settles to a lower final value. Comparison shows that,
in all cases, the final value for the model in Figure \ref{f10} is
slightly higher than the mean in Figure \ref{f8}. Presumably this is
because the mean value of the accreted positive angular momentum is
much greater than the mean negative value.

Finally, we can use Figure \ref{f10} and the first equality of
Equation (5) to get another estimate of the magnitude of energetic
outbursts from the planet. The typical timestep of the numerical model
is about 1.5 days, and from Figure \ref{f10}, the mass of infall onto
the companion is typically a burst of about 30 particles in the $30\
and\ 50\ M_J$ models, respectively. These values translate into a mass
accretion value of $4.6 \times 10^{17}\ g/s$, and a luminosity of
about $0.072\ L_{\odot}$ for the $30\ M_J$ model. The luminosity would
be a few times greater for the $50\ M_J$ model and mass infalls can
also sometimes be a few times greater. Moreover, since the free-fall
time at the mean disc radius is only a few hours, the numerical models
may overestimate the infall time of a mass cluster, and thus,
underestimate the luminosity. Luminosities of order $1.0\ L_{\odot}$
seem quite possible. Note, in particular, that the estimates of
Equation (2) are based on a relatively low Mira mass loss rate ($2
\times 10^{-7}\ M_{\odot}/yr$) and the corresponding mean ambient
density.

As discussed above infall events come in clusters, so we would expect
an outburst to be sustained for times of several days to several
weeks, as seen in the Hipparcos data. 

\subsection{SiO Masers}

In previous papers (Struck-Marcell 1988, Willson and Struck 2001,
Paper I) we noted that the region of SiO maser emission in Miras with
such emission coincides with the pulsating region of the extended
atmosphere. Those works also discussed how the presence of a giant
planet or brown dwarf companion at similar orbital radii could effect
the SiO emission, e.g,, with clumping due to turbulence in the wake to
produce density enhancements needed for the masing, and with
relatively strong local magnetic fields around the companion and in
the wake to help explain the high polarizations of some maser stars.

The new models provide more detail on wake structure, especially in
the case of high mass companions. On the other hand, information on
the structure of clumps is limited, because that will depend on
detailed cooling and heating processes not included in our isothermal
equation of state. Despite this limitation, Figure \ref{f7} still
shows the presence of substantial inhomogeneities in the wakes. The
vortices in Figure \ref{f9} provide an even more striking example of
inhomogeneous flow. The wake bounding shocks visible in Figure
\ref{f9}, combined with the AU length of the wakes in Figure \ref{f7},
show that there are substantial shock surfaces. Wake and pulsational
shocks are of comparable amplitude in the models above. Thus, the wake
environment should be just as favorable as that of the pulsational
shocks for exciting SiO molecules.

Moreover, the last three panels of Figure \ref{f1} show that the
companion and the wake can noticably ripple the outer edge of
pulsational shocks, and perturb the gas behind that edge, over a large
azimuthal angle. As a result, wakes may also enhance clumping behind
pulsational shocks. \citet{har97} have suggested that pulsational
shocks are mediated by strong magnetic fields (of order 50 G), and
that clumps result from the action of the Parker instability in those
shocks. One obvious difference between the two models is that
companion induced clumping occurs around an orbital ring, which could
be highly flattened in some views. The Hartquist and Dyson clumping,
like clumping in the pulsational shocks, would be spherically
symmetric, in the absence of rotational flattening, and probably be
seen as always circular, with limb brightening. The maser spots of
{\it{o Ceti}} (Mira) appear to have a fairly flattened distribution,
which could be a ring viewed edge-on \citep{phi00}.

The range of sizes in the wake and wake inhomogeneities in the
substellar companion model (i.e., of order tenths of an AU up to a few
AU) also seem in accord with VLBI observations of SiO maser spots. For
example, masers spots subtending an angle of 3 milliarcseconds, near a
Mira at a distance of about 300 pc, have a physical extent of about 1
AU. (These numbers derive from \citet{kem97} study of TX Cam, but are
also appropriate for {\it{o Ceti}}, see \citet{phi00}.)

The shortest time-scales reported for variations in SiO maser line
profiles are about 10-20 days \citep{pij94}. This corresponds to a
distance of a few hundredths of an AU for a roughly sonic velocity of
about 5.0 km/s, and still less than 0.1 AU for shocks propagating at
10 km/s. Thus, such rapid changes must be associated with quite small
scale dynamical phenomena, such as a shock passage, collisions between
small clumps, or interactions on the small scale near a companion. On
the other hand, lifetimes of specific spectral features can range up
to several hundred days, which would correspond to a sonic traversal
distance of close to an AU. Such motions could include either
traversal of a companion wake or the development of a clump in a
hydromagnetic instability.

\subsection{More Speculative Possibilities}

In this subsection we consider some more speculative possibilities for
detecting planetary or brown dwarf companions to Miras. These include
effects related to the presence of: 1) a hot corona around the
companion, and 2) intermittent jets launched from the companion.

\subsubsection{Coronae}

We have described above how Mira wind material may fall onto the
companion and an accretion disc with a typical radius of several times
that of the companion radius, at about free fall velocities. With
the companion mass range considered in this paper, these velocities
are of order 300 km/s. Collisions between gas elements with such
relative velocities could generate gas temperatures of millions of
Kelvins in at least a fraction of the gas. At the high densities of
the companion surface, and even the accretion disc surface, cooling is
probably very efficient. However, less dense gas in the atmospheres of
the companion and the disc can be heated either in the collision or by
absorbing the cooling radiation. 

In Paper I we estimated that, if converted to ionizing photons, the
accretion luminosity could ionize a sphere of radius of order $10\
R_J$ around the companion (depending on many poorly known
parameters). The pressure scale height of gas with a temperature of a
few million degrees is also of order $10 R_J$. Thus, this material is
deep within the capture radius of the companion, so any passing
circumstellar gas will be captured, and will not sweep the corona out.

We would expect the coronal gas to be of quite low density relative to
its surroundings. For example, we can require that its free cooling
time be greater than about 1 yr., a typical time between accretion
bursts or pulsational wave passages.  Then with a high temperature
cooling rate of about $2 \times 10^{-23}\ ergs\ cm^{3}\ s^{-1}$ (e.g.,
\citealt{spi78}), we find that the coronal density must be less than
about $1000\ cm^{-3}$. Given this density and a volume corresponding
to the radius above, we find that the emission associated with coronal
cooling is negligible ($<\ 10^{-17}\ L_{\odot}$). Generally, we expect
that to obtain a detectable coronal emission measure would require a
material density that is high enough to give a cooling time of much
less than a year. Thus, while a corona is an interesting component of
the companion region, is not directly observable. Most of the
accretion energy will be emitted from transient hot spots on the disc
and companion.

\subsubsection{Magnetosphere and Coronal Wind}

The companion may have a magnetic field, and magnetosphere, like that
of Jupiter. If so, it would have an important effect on the corona and
the accretion disc. First of all, we note that the magnetic pressure,
given a surface field of about 3.0 G, would be of the order,

\begin{equation}
P_B = 3.6\ \times \ 10^{-7} {\biggl(}\frac{B_{surf}}{3.0G}{\biggr)}^2
{\biggr(}\frac{10R_J}{R}{\biggl)}^6\ dyne\ cm^{-2},
\end{equation} 
at a distance R from the companion centre. This is generally
comparable to or greater than the thermal pressure of a maximum
density corona given by, 

\begin{equation}
P_{th} \le 6.2\ \times \ 10^{-7} {\biggl(}\frac{n}{1000\
cm^{-3}}{\biggr)} {\biggl(}\frac{T}{3.0 \times 10^6 K}{\biggr)}. 
\end{equation}
When hot gas is created in accretion impacts it will begin to expand
out into the corona, but it will also be constrained by the magnetic
field. 

If the companion and its magnetosphere are rotating like Jupiter, then
the centrifugal force will be important, e.g., in driving a
centrifugal wind. \citet{sok00} and \citet{sok02} have also discussed
the formation of companion winds around AGB stars, However, the case
they considered was that of a white dwarf or main sequence companion,
which is able to spin up the Mira atmosphere, and induce strong Roche
lobe accretion onto the companion. This induces a strong fast
wind. The cases considered in this paper generally fail to meet Soker
\& Rappaport's condition for strong tidal interaction, though the most
massive brown dwarfs orbiting within a few AU would be
affected. Therefore we would not expect as strong a wind as described
in their work.

Nonetheless, once the accretion disc builds up to a substantial mass,
the enviroment around it my be quite similar to that around a magnetic
protostar. The X-wind theory of Shu and collaborators (e.g.,
\citet{shu99} and references therein) may be applicable. However, the
mass and energy in such a variable wind would be much less than in the
protostellar case, so it seems unlikely that these effects would be
directly observable. 

\section{Summary and Conclusions}

In the models described above we have studied the interaction between
the extended, pulsating atmosphere and wind of a Mira variable and a
brown dwarf or giant planet companion. Our numerical models resolved
the time-dependent bow shock and wake structure created by the
companion. The outermost parts of the accretion onto the companion
were also resolved. We obtained the following results concerning the
gas flow.

1) The accretion rate scales roughly linearly (in two dimensions) with
the companion mass. (See Fig. \ref{f3} and Section 3.2.)

2) The equation of state, or more precisely the value of the exponent
$\gamma$ of the adopted barotropic equation, has little effect on the
accretion rate. Naively, this is because accretion is the result of
gravitational capture under conditions where the thermal pressure is
small.

3) One of the most interesting results above is the transition from
periodic to chaotic mass accretion, and from chaotic to periodic
angular momentum accretion as the companion mass is increased through
the range considered in this paper ($10-50\ M_J$). (See Figs. \ref{f4}
and \ref{f8}, and Sections 3.3 and 3.4.)

4) The wakes grow substantially as the companion mass is increased
through this range. Wake structure, and in particular, clumping could
have an important effect on SiO maser emission, which derives from the
same region. (See Fig. \ref{f6} and Section 4.3).

While we do not resolve the flow on the inner scale ($R < 200\ R_J$),
we did monitor the mass and angular momentum accreting into this
region. With these data and a few simplifying assumptions, we examined
the properties of a simple semi-analytic model for this flow (Section
4.2). Using this model we obtained the following additional results
for flow on the inner scale.

5) The effective boundary conditions between the inner and outer
scales are important. In Paper I we adopted a boundary condition
equivalent to the assumption that it was difficult for accreting gas
to penetrate to the inner scale. This lead to strong bursts of
accretion. This case requires coronal or magnetospheric pressure to be
strong enough to resist the infalling gas. Estimates of the corona
derived from our semi-analytic model, and of magnetospheres comparable
to that of Jupiter, suggest that that would not be the case with
companions like those studied in this paper.

6) The semi-analytic model suggests that an accretion disc would form
around the companion on a scale of order a few companion radii (see
Section 4.4).

7) The pattern of angular momentum accretion (positive and negative)
suggests that accretion bursts are likely. They may be strong enough
to generate observable effects in the optical (see Section 4.1).

8) Energy dissipation from the impact of free-falling gas elements on
the companion and the accretion disc may generate a thin hot corona
around the companion, and possibly a coronal wind (see Section
4.4). It is unlikely that these would be directly observable.

In sum, the situation studied in this paper is a previously unstudied
type of close binary interaction. This interaction is different from
that of the more well known symbiotic and cataclysmic type binaries in
several respects. First of all, the close companion in those systems
is generally a more massive white dwarf or M dwarf, and the
interaction may involve mass transfer via Roche lobe overflow, or
interacting winds, rather than direct accretion out of a Mira wind. A
second aspect of the present study is the inclusion of a radially
oscillating environment, which we found has profound effects on the
accretion process.

The oscillating environment has not generally been considered in
models of other types of close AGB binaries. However, as \citet{mas98}
and \citet{sok02} have pointed out, a close massive companion would
exert significant tidal forces on the Mira atmosphere. Moreover, the
Mira would have a significant motion around the centre of mass. These
and other relatively energetic phenomena make it unlikely that such
systems would be classified as ordinary Miras. Such cases deserve
further study.

Several other issues raised above also merit further study. The first
is the thermohydrodynamics of the inner scale. Fully
self-consistent numerical modeling with heating and cooling and
magnetic fields included would be very interesting, if not entirely
feasible at present. Including heating, cooling and magnetic effects
in the wake flow would also be very interesting, especially to see the
impact of these processes on SiO maser emission. However, the most
urgent need is for more observational constraints on the models. The
models above provide a number of observationally testable
predictions. The observations published to date provide intriguing
hints, but are not sufficient to provide strong evidence for the
existence of such systems or strong tests of the models.

\section*{Acknowledgments}

We are grateful to G. H. Bowen for many helpful conversations on
Miras, and to an anonymous referee for numerous helpful
suggestions. This research has made extensive use of the NASA IPAC
Extragalactic Database (NED).

\label{lastpage}

\end{document}